\documentclass[twocolumn,showpacs,amsmath,amssymb]{revtex4}

\usepackage{graphicx}
\usepackage{dcolumn}
\usepackage{bm}
\newcommand{\etal}{\emph{et al.}}
\newcommand{\be}{\begin{equation}}
\newcommand{\ee}{\end{equation}}
\newcommand{\bfig}{\begin{figure}}
\newcommand{\efig}{\end{figure}}

\begin{document}
\title{Tuning the quantum oscillations of surface Dirac electrons in the topological
insulator Bi$_2$Te$_2$Se by liquid gating.
}
\author{Jun Xiong$^1$, Yuehaw Khoo$^1$, Shuang Jia$^2$, R. J. Cava$^2$ and N. P. Ong$^1$
}
\affiliation{
Department of Physics$^1$ and Department of Chemistry$^2$, Princeton University, Princeton, NJ 08544
}

\date{\today}
\begin{abstract} 
In Bi$_2$Te$_2$Se, the period of quantum oscillations arising from surface Dirac fermions
can be increased 6-fold using ionic liquid gating. 
At large gate voltages, the Fermi energy reaches the $N$ = 1 Landau level in a 14-Tesla field. This 
enables the $\frac12$-shift predicted for the Dirac spectrum to be measured accurately.
A surprising result is that liquid gating strongly enhances the surface mobility. 
By analyzing the Hall conductivity, we show that the enhancement occurs on only one surface.
We present evidence that the gating process is fully reversible (hence consistent with 
band-bending by the $E$-field from the anion layer accumulated). In addition to the surface carriers, the experiment 
yields the mobility and density of the bulk carriers in the impurity band. 
By analyzing the charge accumulation vs. gate voltage, we also obtain
estimates of the depletion width and the areal depletion capacitance $C_d/A$. The value of $C_d/A$ implies 
an enhanced electronic polarizability in the depletion region.
\end{abstract}


\maketitle                   
\section{Introduction}
A Topological Insulator (TI) is characterized by the existence of current-carrying surface states that
traverse the bulk energy gap~\cite{KaneHasan,QiZhang}. 
There is strong interest in the helical nature of the surface states,
which results from the locking of the electron's spin transverse to its momentum. 
In the bismuth-based TI materials, photoemission 
spectroscopy~\cite{Hsieh} and scanning tunneling microscopy (STM)~\cite{Roushan} 
have confirmed the spin-locking feature. In transport experiments, the surface states have 
been detected by surface Shubnikov
de Haas (SdH) oscillations in Bi$_2$Te$_3$~\cite{Qu} and (Bi,Sb)$_2$Se$_3$~\cite{Fisher}, and 
by Aharonov-Bohm oscillations in Bi$_2$Se$_3$ nanowires~\cite{AB}.

Among the Bi-based TI materials, Bi$_2$Te$_2$Se currently displays the highest bulk 
resistivities ($\rho = \;1-6\;\Omega$cm at 4 K)
~\cite{Ando10,Xiong12a,Jia,Xiong12b}. Despite the large $\rho$,
SdH oscillations may be tracked to temperatures $T$ as high as 38 K~\cite{Xiong12a}.
A persistent problem, however, is that the surface Fermi energy $E_F$ in as-grown crystals is still 
quite high ($\sim$200 meV above the Dirac Point). An \emph{in-situ} method that demonstrably tunes $E_F$
would greatly facilitate experiments at the Dirac Point, as proposed in Refs.~\cite{KaneHasan,QiZhang}.
Several groups have applied conventional electrostatic gating to tune the chemical potential $\mu$ in
exfoliated crystals~\cite{Check11,Pablo,Morpurgo} and in thin-film samples of Bi$_2$Se$_3$.~\cite{Steinberg} 
The newer technique of liquid gating has also been used on Bi-based materials~\cite{Iwasa,Fuhrer,Check12,Iwasa12,Ando12}.
However, in these experiments, SdH oscillations were either not detected at all or poorly resolved~\cite{Morpurgo} even if the
second derivative is used.
Tuning of the quantum oscillations and showing that they arise from surface Dirac electrons
remain to be established.

\begin{figure}[t]
\includegraphics[width=7 cm]{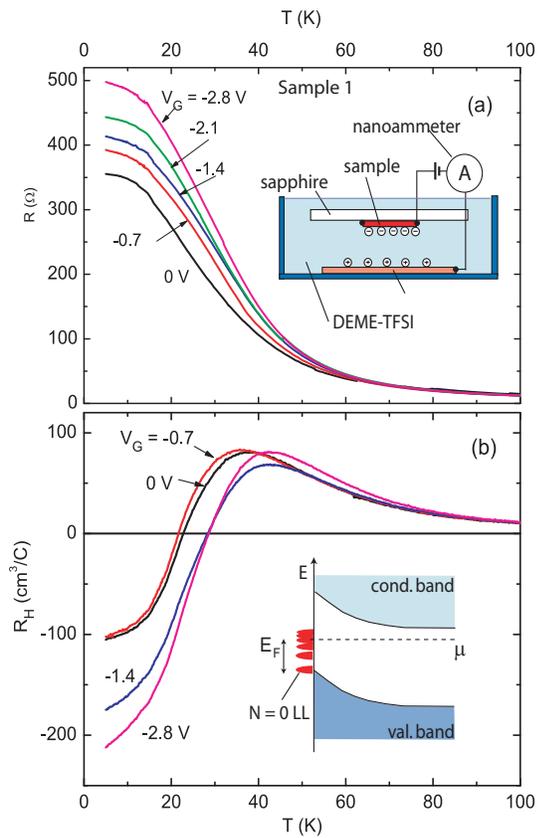}
\caption{\label{figRRH} (color online) The resistance $R$ per square (Panel a) and
Hall coefficient $R_H$ vs. $T$ (Panel b) in Bi$_2$Te$_2$Se at selected $V_G$ in Sample 1. 
$R_H$ is measured at fixed $B$ (3 T). Changing $V_G$ from 0 to -2.8 V increases $R$ 
by 40$\%$ and $|R_H|$ by 2$\times$. The inset (Panel a) shows 
the cell housing the sample and the ionic liquid DEME-TFSI. The Au electrode (a
circular plate of radius 1.5 mm) is separated by 0.5 mm from the sample.~\cite{liquid}
The inset in (b) is a sketch of the band bending induced by liquid gating. 
Negative ions deposited on the crystal leads to upward band-bending. At the surface, this causes
$E_F$ to decrease towards the Dirac Point. LLs are shown as solid half-ovals.
}
\end{figure}


Here we report that the surface SdH oscillations in Bi$_2$Te$_2$Se can be tuned over a broad range using
the ionic liquid (DEME-TFSI). In lowering $E_F$ substantially, we access the $N$ = 1 Landau
level in a magnetic field $B$ = 14 T. This allows the $\frac12$-shift characteristic of
Dirac electrons to be measured with greatly improved resolution. We find that the intercept 
remains fixed at $\frac12$ even as the surface density is tuned by factors of 3-6.
An unexpected finding is 
that liquid gating leads to strong enhancement of the mobility  
$\mu_s$ of the surface carriers. We attribute the enhancement to the ``smoothing'' of local potential
fluctuations seen by the Dirac fermions. Aside from moving $E_F$ closer to the Dirac Point, the
tunability yields direct information on the surface and bulk conduction. The additional information
enables us to determine how the surface and bulk mobilities change with gate voltage $V_G$.
We discuss the evidence that the liquid gating in our experiment is causing band bending rather
than unwanted chemical reaction. Lastly, thanks to the SdH oscillations, we can measure 5 parameters at each setting of $V_G$
($E_F$ and $\mu_s$ of the surface carriers, the bulk density and mobility, and the total ionic
charge $Q$ deposited). The 5 parameters provide cross checks for the gating experiment. In particular, 
we determine the depletion capacitance $C_d$ which measures the polarizability of the depletion region.

As reported earlier~\cite{Ando10,Xiong12a,Xiong12b}, the resistance $R(T)$ in Bi$_2$Te$_2$Se rises  
monotonically to very large values as $T\to$ 4 K (curve at $V_G$ = 0 in Fig. 
\ref{figRRH}a). Analysis of the Hall coefficient $R_H$ at 5 K (Fig. \ref{figRRH}b) 
reveals a population of bulk $n$-type carriers much higher than the population of surface electrons. 
Nonetheless, a modest, negative gate voltage $V_G$ can increase $R$ by 40$\%$ (Fig. \ref{figRRH}a)
and $|R_H|$ by a factor of 2 at 5 K (Panel b). $V_G$ is applied to the gold electrode (inset in Fig. \ref{figRRH}a) at 220 K, and the sample is then cooled below the liquid's glass transition. After the low-$T$ measurements are
completed, the sample is warmed to 220 K (at 2 K/min.) and $V_G$ is reset.
The ``gating'' temperature is selected within the optimal window 220-240 K (see below and Appendix \ref{apptemp}).
At 4 K, the large $E$-field induced by the surface anion charge $Q$
(1-4$\times 10^{14}e$ cm$^{-2}$) creates a depletion layer that penetrates deep into the bulk (5-20 $\mu$m),
where $e$ is the electron charge. As shown in Fig. \ref{figRRH}b (inset), the induced upward bending of the bands decreases $E_F$.

At each value of $V_G$, the curves of $R$ vs. $B$ display SdH oscillations. 
To focus on the SdH signal, we 
have subtracted off a smooth background $\rho_B$ to isolate the oscillatory part 
of the resistance, $\Delta \rho_{xx} \equiv \rho_{xx} - \rho_{B}$.
Figure \ref{figSdH}a displays plots of $\Delta\rho_{xx}$ in Sample 1
versus $1/B$ for 5 values of $V_G$. The period of the SdH oscillations increases
monotonically as $V_G$ changes from 0 to -4.2 V, in accordance with 
our expectation that $E_F$ is decreasing.
Surprisingly, the SdH amplitude is strongly enhanced between $V_G = 0$ and -2.1 V 
(the former is shown amplified by 5$\times$). 
The dotted curves are the best fits ~\cite{Xiong12b} to the
Lifshitz-Kosevich (LK) expression for SdH oscillations using only one 
frequency component. From the fits, we may infer how the surface mobility $\mu_s$ 
changes with $V_G$ (see below).
The same trends are evident in 
Sample 2, which has a higher 
starting surface density $n_s$ but is taken to $B$ = 45 T (Fig. \ref{figSdH}b). 
We find that the SdH oscillations are not resolved at $V_G=0$, but
become prominent at $V_G$ = -1.5 V.

\section{Experimental details}
In our experiment, the sample is immersed in the 
ionic liquid DEME-TFSI, comprised of 
cations (CH$_3$CH$_2$)$_2$(CH$_2$CH$_2$OCH$_3$)CH$_3$N$^+$ and anions (CF$_3$SO$_2$)$_2$N$^-$.
The liquid is pumped at 25$^o$C for 2 hours prior to application to minimize water content.
Liquid gating has several pitfalls when used on crystals. 
After the gate voltage $V_G$ is selected at a ``gating temperature'' (Appendix \ref{apptemp}),
the sample is cooled to 5 K for the SdH measurements.
The stresses induced by repeated freezing and thawing of the ionic liquid can
snap the leads or the crystal itself. Also, a large $|V_G|$ can trigger
an electrical discharge which invariably leads to a steep collapse of $R$ (at 5 K). 
Unlike in thin films, changes to $R$ with $V_G$
are not resolved above $\sim$100 K (see Fig. \ref{figRRH}a). 
To minimize sample damage, we start at $V_G = 0$ 
followed by measurements at increasingly negative $V_G$ until the sample fails (usually by a discharge event). 
At 220 K, $V_G$ is changed in steps of -0.1 V, while monitoring the transient current $I_{trans}$ (1-40 nA). 
The time spent at 220 K is typically 300-500 s. We emphasize that the changes to
$\rho$ and $n_H$ are reversible (see below).
On returning $V_G$ to 0, we recover the same starting value of $R$ (at 5 K)
provided $|V_G|$ is kept below $\sim$2 V.

\begin{figure}[t]
\includegraphics[width=9 cm]{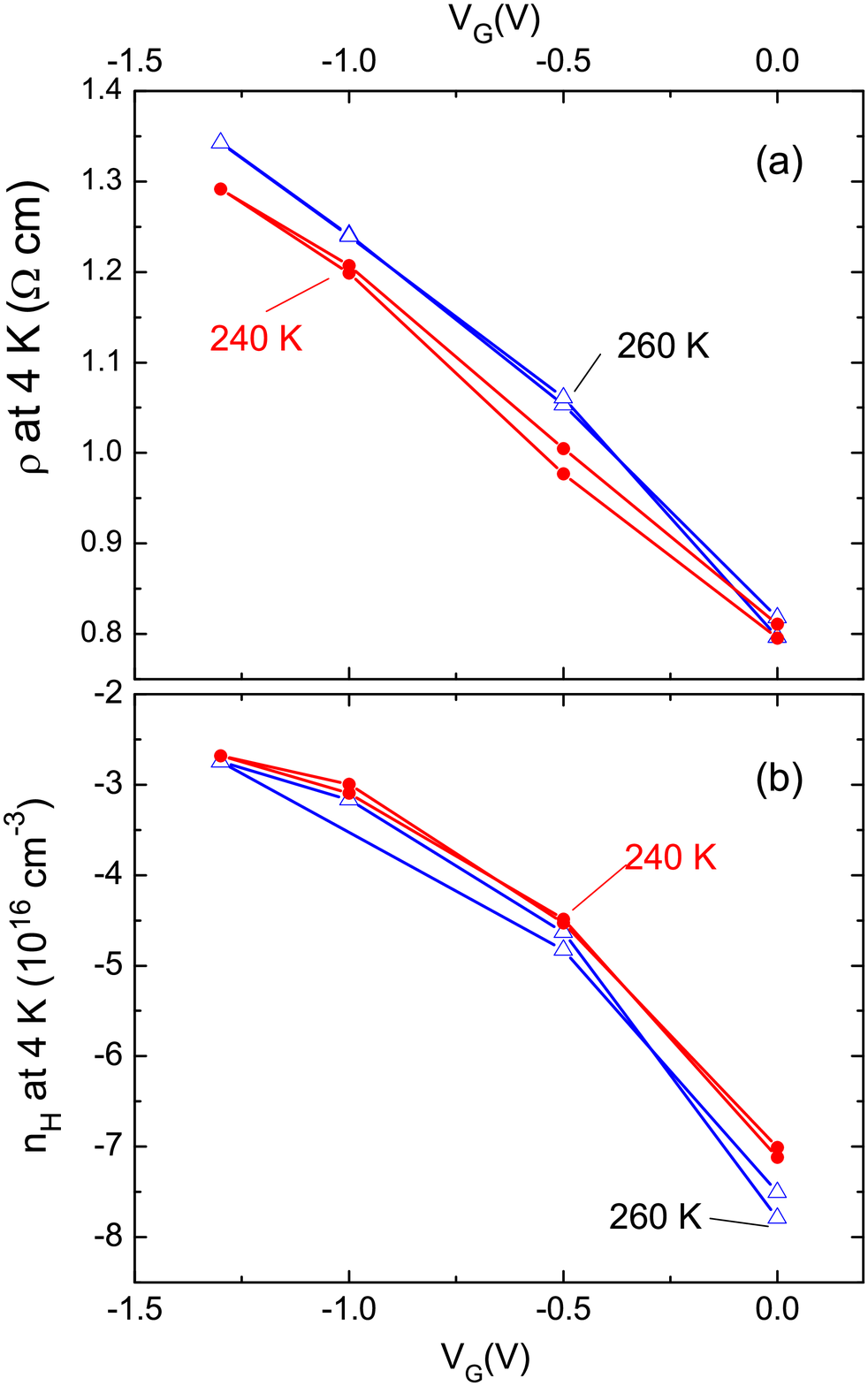}
\caption{\label{figtest} (color online) Test experiments to show negligible hysteresis in the
sample's resistivity $\rho$ (Panel a) and Hall density $n_H$ (Panel b), as $V_G$ is changed from 0 to -1.3 V, then back to 0 V
at the temperatures $T$ = 240 and 260 K (Sample 3). The small hysteresis (within the measurement uncertainties) is taken as
evidence that chemical reaction is negligible compared with the physical gating effect.
The accumulation time is 800 s.
}
\efig

The crystal dimensions of Sample 1 are $0.9 \times 0.75 \times 0.05$ mm$^3$. For Sample 2, they are 1.35$\times$0.61$\times$0.026 mm$^3$. 
In Sample 2, the steepest change in $\mu_s$ occurs between $V_G = 0$ and
-1.5 V, at which $\mu_s$ = 2,800 cm$^2$/Vs. At larger gate, it saturates
($\mu_s$ = 3,000 cm$^2$/Vs at -6 V). 

The possibility that the
strong $E$-field can induce chemical doping of the sample
is an important concern in liquid-gating experiments. We note that the effects of chemical 
reaction are inherently non-reversible. In particular, let us assume that chemically induced doping
occurs at some finite $V_G$ leading to changes in $\rho$ and $n_H$ (measured when cooled to 4 K). When
the sample is rewarmed to 240 K and $V_G$ is reset to zero, we should not expect $\rho$ and $n_H$ to
recover their starting values when recooled to 4 K (resetting $V_G$ to zero cannot reverse
the chemical damage). Hence, 
we adopt the working assumption that the absence of resolvable hystereses in $\rho$ and $n_H$ (measured
at 4 K) as $V_G$ is cycled provides evidence that band-bending is the dominant effect and chemical
reaction effects are minimal. We have performed a much broader set of tests (on Sample 3) to 
investigate details of the ion-accumulation process over an extended range of gating 
temperatures (208 $<T<$ 260 K). Samples 1 and 2, from which the detailed SdH results were obtained, were
not subject to these cycling processes to minimize stress damage.

Figure \ref{figtest} shows the variation of $\rho$ (Panel a) and $n_H$ in Sample 3
as $V_G$ is changed step-wise from 0 to -1.3 V and back. 
After the $V_G$ is set anew (at the gating $T$ = 240 K), we wait for 800 s
to accumulate the anions before cooling to 4 K for the measurements of $\rho$ and $n_H$.
By monitoring the transient charging current $I_{trans}$, we have also measured the
ion accumulation charge $Q(t)$ (Appendix \ref{appcap}).
The absence of hysteresis, within the experimental uncertainty, is evidence
that the changes are fully reversible and hence caused by band-bending effects. 
The experiment, repeated at 260 K, also shows negligible hysteresis.

Apart from chemical reaction, two other important factors are incomplete melting of the
ionic charge configuration when $T$ is too close to the glass transition and the intrinsic
(activated) bulk conductance of the ionic liquid. These additional factors have been
investigated using the measured $Q(t)$. We discuss them in 
Appendix \ref{apptemp}.

\section{Tuning the surface Dirac State density}\label{Tuning}

\begin{figure}[t]
\includegraphics[width=9.5 cm]{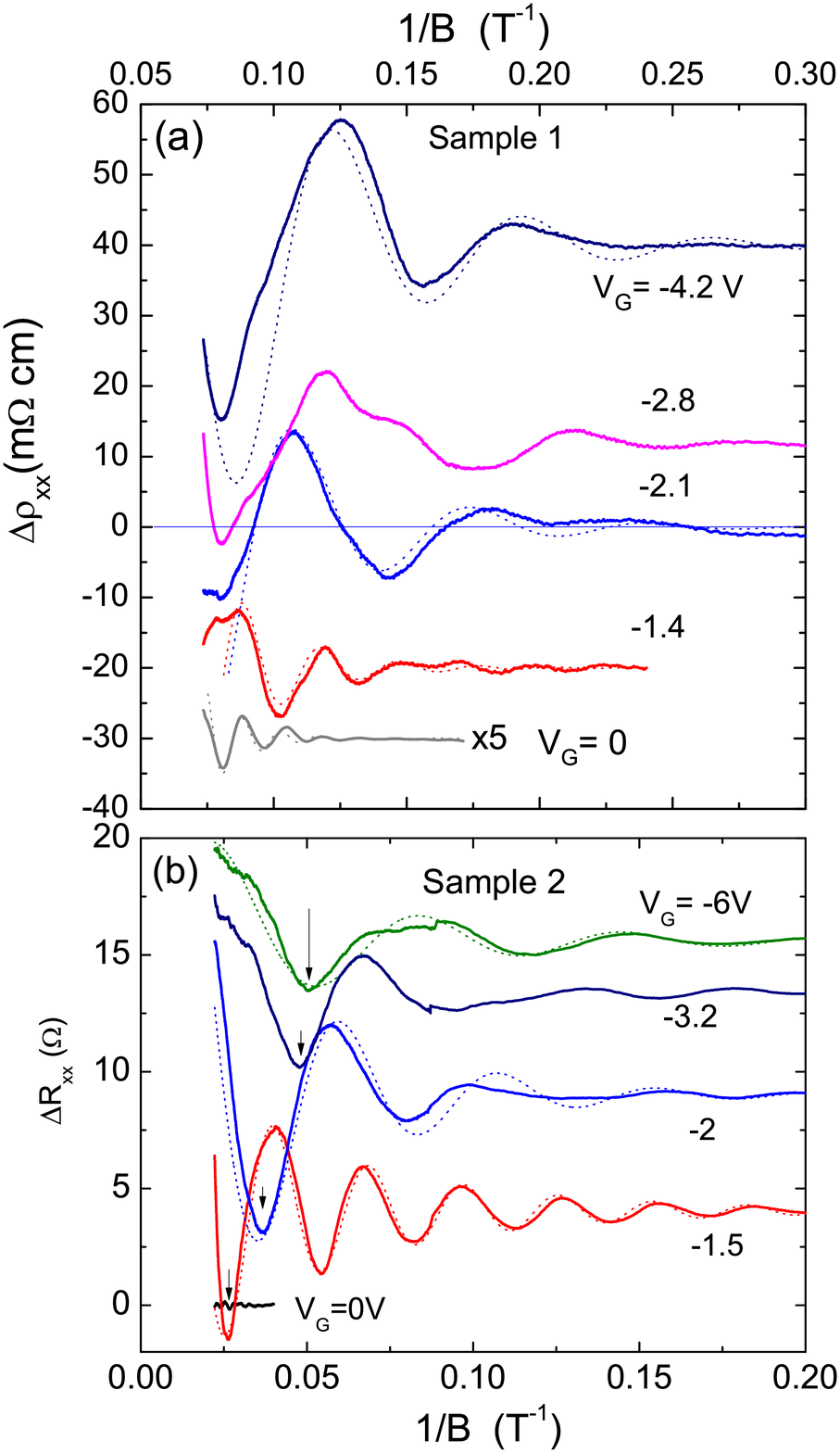}
\caption{\label{figSdH} (color online)
Traces of SdH oscillations in the resistance versus $1/B$,
showing systematic changes to the oscillation amplitude and period with gate voltage
(bold curves, displaced vertically for clarity). The dashed curves
are fits to the LK expression with one frequency component~\cite{Xiong12b}. 
Panel (a) shows traces of $\Delta\rho_{xx}$ 
vs. $1/B$ at 5 K measured to 14 T for 5 values of $V_G$ (Sample 1). 
The largest increase in amplitude occurs between $V_G$ = 0 to -1.4 V. The curve
at $V_G = 0$ is shown amplified $5\times$. All other curves share 
the same vertical scale. Panel (b) displays traces of $\Delta R_{xx}$ vs. $1/B$ at 1.6 K 
measured to 45 T at $V_G$ as indicated (Sample 2).
Arrows indicate $n=\frac12$ ($E_F$ at center of broadened $N$ = 1 LL). 
}
\end{figure}

\begin{figure}[t]
\includegraphics[width=9.5 cm]{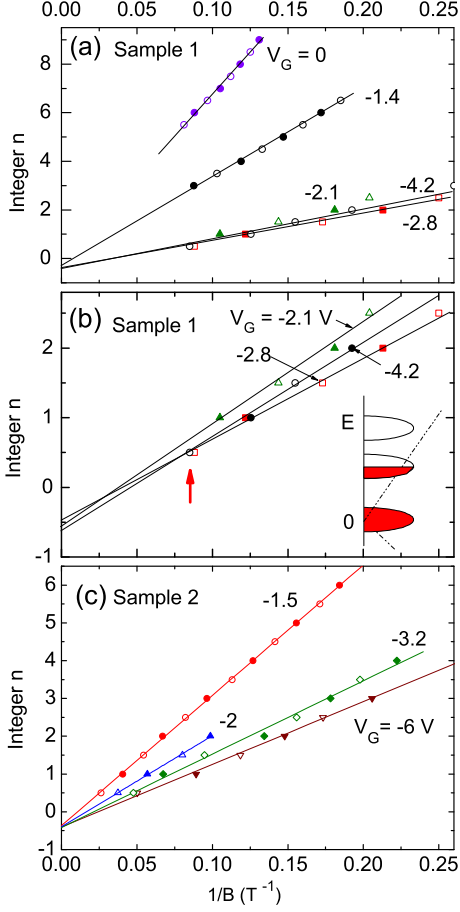}
\caption{\label{figIndex} (color online)
Index plots of the integer $n$ vs. $1/B_n$ at selected $V_G$ in Sample 1 (Panels a and b) and 2 (Panel c).
Maxima of $\Delta\rho_{xx}$ (solid symbols), corresponding to the index fields $B_n$, are plotted against $n$. Minima (open symbols)
are plotted against $n+\frac12$.
Panel (a): As $V_G$ changes from 0 to -2.1 V, the slope of the best fit lines decreases 6-fold. Further increase
in $|V_G|$ leads to saturation. In Panel (b), the high-bias curves are displayed in expanded vertical scale. In
the limit $1/B\to 0$, the best-fit lines have intercepts at -0.46, -0.56 and -0.61, consistent with Dirac fermions.
The datum at $n=-\frac12$ (arrow) corresponds to $E_F$ sitting in the middle of the broadened $N$ = 1 LL (inset).
The intercepts for Sample 2 (Panel c) also cluster near -0.45 in the limit $1/B\to 0$.
}
\end{figure}

\begin{figure}[t]
\includegraphics[width=9.5 cm]{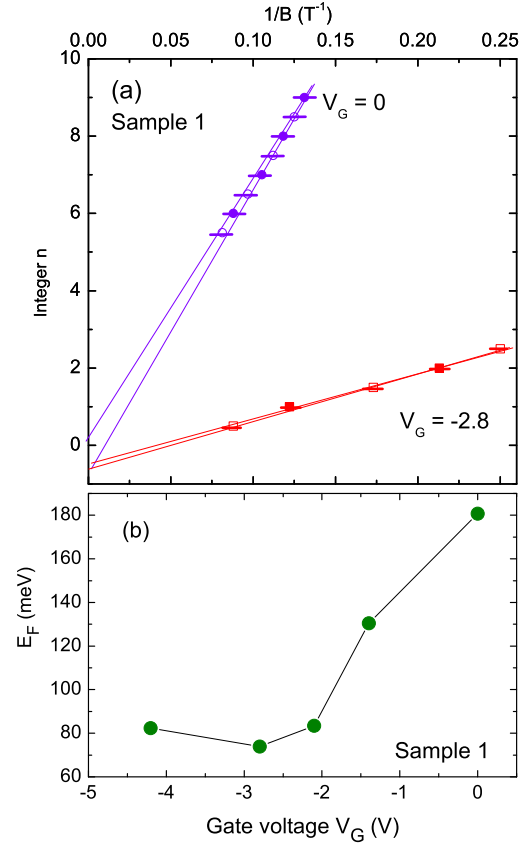}
\caption{\label{figError} (color online)
(Panel a) Comparisons of extrapolations of index plot to the limit $1/B\to 0$ for index fields measured
with $V_G=0$ (circles) and measured with $V_G$ = -2.8 V (squares) in Sample 1.
(Panel b) The variation of $E_F$ with applied $V_G$ in Sample 1 ($E_F$ is measured from the Dirac Point). 
The FS cross-section $S_F$ is converted to $E_F$ by $E_F = \hbar v\sqrt{S_F/\pi}$ 
using $v$ = 6$\times 10^5$ m/s~\cite{Xiong12a}. For $|V_G|> 2 V$, the decrease in $E_F$ saturates.
}
\end{figure}

\begin{figure}[t]
\includegraphics[width=9 cm]{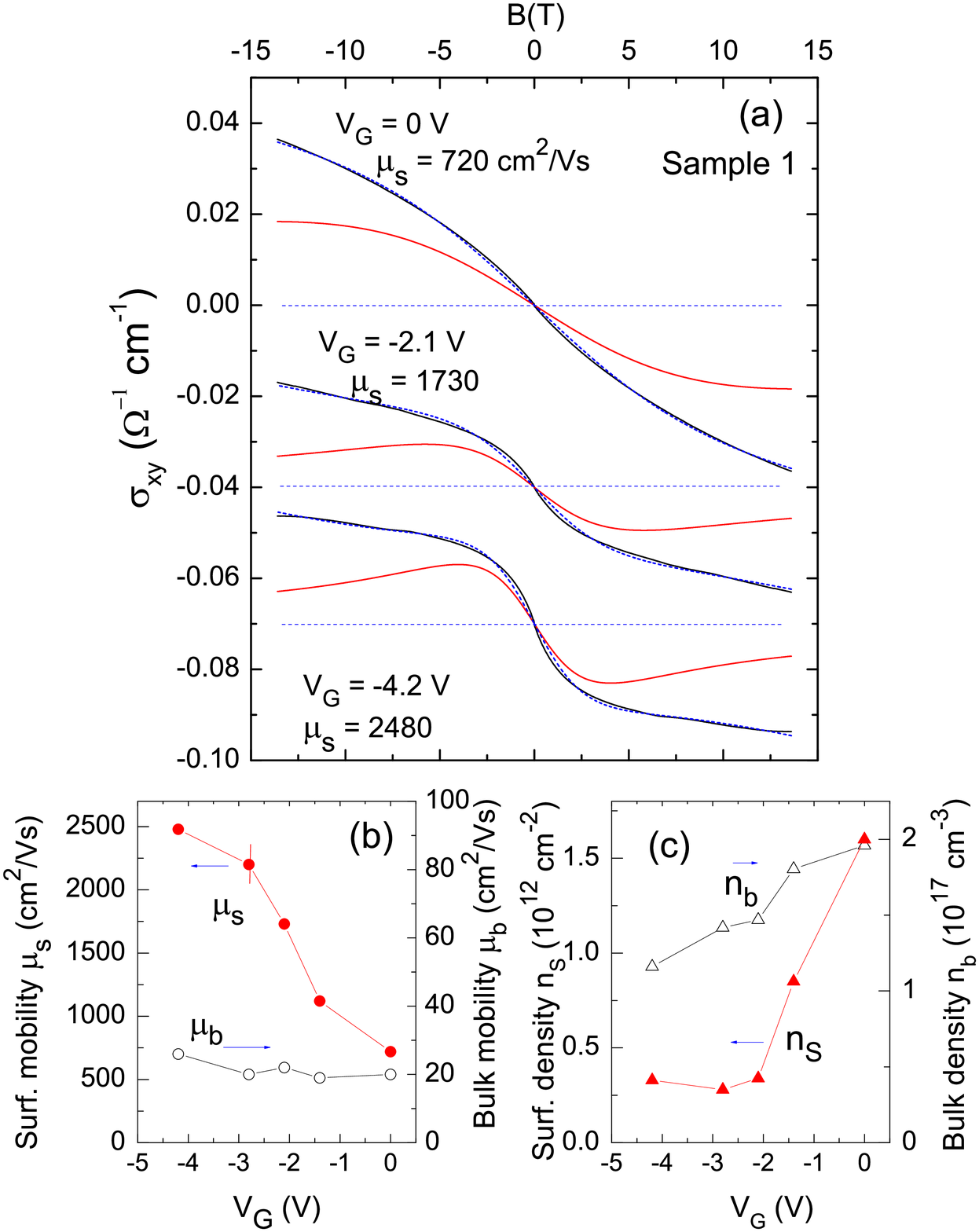}
\caption{\label{figHall} (color online) 
Panel (a): The observed Hall conductivity $\sigma_{xy}$ vs. $B$ in Sample 1,
showing weak-$B$ curvature at 3 values of $V_G$
(curves displaced for clarity). At each $V_G$, the outer curves are the 
data (solid black curve) and the fit to Eq. \ref{eq:sxy} (superposed blue dashed curve). The inner
(red, solid) curve is the surface term
$G^s_{xy}/t$ fixed by $n_s$ and $\mu_s$. The difference 
between the outer and inner curves is the bulk term $\sigma^b_{xy}$. 
At $V_G$ = -4.2 V, $G^s_{xy}/t$ accounts for 83$\%$ of
$\sigma_{xy}$ in weak $B$. 
Panel (b) shows that, with increased gating, $\mu_s$ increases from 720 to 2,480 cm$^2$/Vs 
while $\mu_b$ stays very small (20-30 cm$^2$/Vs). Panel (c) 
compares the sharp decrease in $n_s$ with the mild change in $n_b$ with gating. When $|V_G|>$
2 V, $n_s$ saturates.
}
\end{figure}

In finite $B$, the surface electronic states are quantized into Landau levels (LLs) 
with quantum numbers $N = 0,1,\cdots$. The index field $B_n$ is the field at which
$E_F$ falls between two LLs.
For Schr\"odinger states, the integer $n$ counts the number of 
occupied LLs (the highest filled LL has $N_{max} = n-1$). Using the level 
degeneracy $Be/h$ per spin, we then have $1/B_n = ne/(hn_s)$ 
($n_s$ is the surface density, $e$ the elemental charge and $h$ is Planck's constant).

For Dirac electrons, however, we have
$n+\frac12$ filled LLs when $B = B_n$ (now $N_{max} = n$).
The additional $\frac12$ derives from the $N=0$ LL, or equivalently, 
from the $\pi$-Berry phase intrinsic to each Dirac cone~\cite{KimStormer}. 
The relation between $1/B_n$ and $n$ is now 
$ 1/B_n = (n+\frac12)(e/h n_s)$ -- a straight line that 
intercepts the $n$-axis at $n = -\frac12$. 
In both cases, $G_{xx}$ is a local minimum at $B_n$.

If resistivity curves are used, $B_n$ should be 
identified with the \emph{maxima} in $\Delta\rho_{xx}$. This point is 
discussed in Refs.~\cite{Xiong12b,Index}
In Fig. \ref{figIndex}a, we plot as solid symbols 
$B_n$ in Sample 1 against the integers $n$ 
(the open symbols corresponding to the minima are plotted against $n+\frac12$).

At each $V_G$, the slope of the straight lines yields the FS area $S_F$. 
As $|V_G|$ increases from 0 to 2.8 V, the slopes of the best-fit lines 
decrease by a factor of 6.4, reflecting a steep decrease in $S_F$. 
This decrease saturates when $|V_G|$ exceeds 2.1 V.

In Fig. \ref{figIndex}b, we show
the high-field behavior for $|V_G|>$ 2.1 V in expanded scale. 
At these large bias values, the intercepts cluster around $n = -\frac12$ 
(-0.46, -0.56 and -0.61). 
In the corresponding plots for Sample 2 (Panel c), we find that $S_F$ decreases by a factor
of 2 between $V_G$ = -1.4 and -6 V. In the limit $1/B\to 0$, the intercepts are at
$n= -0.35$, -0.40 and -0.42. 
In both samples, the last feature observed at the highest $B$ (minima in $R_{xx}$) 
corresponds to $n_{min}=\frac12$
(as shown in the inset of Panel (b), this implies that $E_F$ lies in the middle of the broadened $N=1$ Dirac LL).
With such a small $n_{min}$, we may rigorously
exclude an intercept at $n=0$ in the limit $1/B\to 0$.~\cite{Index}
Thus the index plots provide rather conclusive evidence that the SdH 
oscillations arise from surface Dirac fermions.

The ability to reach $n=\frac12$ is important for experimentally determining the Berry phase $\pi$-shift.
Because the surface SdH oscillations are generally very weak (for $B<$14 T), there is considerable
uncertainty in determining the intercept in the limit $1/B\to 0$ if the
lowest $n$ achieved at the maximum available $B$ is 5 or larger.
We illustrate the uncertainties incurred in Fig. \ref{figError}a. In the absence of gating ($V_G=0$),
the uncertainties $\delta B_n$ in measuring the index fields are typically $\pm 5\%$ (circles).
As shown, this yields a considerable spread in the allowed intercepts (the lowest datum corresponds to $n=5.5$).
For comparison, at the gate voltage $V_G$ = -2.8 V (squares), the lowest datum corresponds to $n=\frac12$. 
This tightens up considerably the spread in the allowed intercepts. The same
advantage may be achieved by applying an intense $B$ (45 T), as was done in Ref. \cite{Xiong12b}.

For convenience, we have converted the values of $S_F$ inferred from the slopes 
of the index plots in Fig. \ref{figHall}a,b to the Fermi energy $E_F = \hbar v\sqrt{S_F/\pi}$
using the Fermi velocity $v$ = 6$\times 10^5$ m/s~\cite{Xiong12a}. 
The variation of $E_F$ vs. $V_G$ in Sample 1 (Fig. \ref{figError}b) shows that $E_F$ 
decreases by 100 mV as $V_G$ is changed from 0 to -2 V. Thereafter, it remains
at $\sim$ 80 mV. This is used below to estimate the depletion capacitance.
We remark that $E_F$ stops decreasing when $|V_G|$ exceeds 2 V. Since the Dirac Point
is close to the top of the valence band, we do not reach the limit of band inversion 
(creating an accumulation layer of holes) in this experiment.

Returning to Fig. \ref{figSdH}, we have fitted the SdH oscillations to the 
LK expressions (shown as dashed curves). The damping of the oscillations versus $B$
yields the surface mobility $\mu_s$. As shown in Fig. \ref{figHall}b,
$\mu_s$ in Sample 1 rises from 720 to 2,480 cm$^2$/Vs as $|V_G|$ is increased to 4.2 V. 
In Fig. \ref{figHall}c, the decrease and eventual saturation in $S_F$ is plotted as a surface density 
$n_s = k_F^2/(4\pi)$ (per spin). The saturation at large $|V_G|$ either arises
from induced chemical reaction or from $E_F$ meeting the top of the valence band.

One figure-of-merit in TI crystals is the ratio of the surface to bulk conductances
$\eta \equiv G^s/G^b$ in zero $B$ (with $G^r \equiv G^r_{xx}(0)$, $r=s,b$).
In Sample 1, $\eta\sim 0.05$ is quite small (compared with $\eta\sim 1$
obtained in Ref.~\cite{Xiong12a}). 
However, in the Hall channel, the ratio $\eta_H = G^s_{xy}/G^b_{xy}$ 
of the surface and bulk conductances ($G^s_{xy}$ and $G^b_{xy}$, respectively)
is enhanced by $\mu_s/\mu_b$, which can be very large. We define $n_b$ and $\mu_b$
to be the bulk electron density and mobility, respectively, averaged over the whole crystal.

As shown in Fig. \ref{figHall}a, a distinctive feature of $\sigma_{xy}$ at low $T$ 
is the curvature in weak $B$, which grows with increasing $|V_G|$. We may use the
semiclassical 2-band expression for $\sigma_{xy}$:
\be
\sigma_{xy}= n_se\mu_s\frac{\mu_sB}{t[1+ (\mu_s B)^2]} + n_be\mu_b^2 B,
\label{eq:sxy}
\ee
where the first term is $G^s_{xy}/t$, with $t$ the thickness (50 $\mu$m in Sample 1). 
With $n_s$ and $\mu_s$ fixed by analysis of the SdH oscillations, this term is 
non-adjustable. The second term is the bulk Hall conductivity
$\sigma^b_{xy}$ in the low-mobility limit.  
With the sole adjustable parameter $P_b\equiv n_b\mu_b^2$, we find that Eq. \ref{eq:sxy} 
gives a very good fit (dashed curves). 
For comparison, we have also plotted $G^s_{xy}/t$ (inner, faint solid curves). 
Combining $P_b$ with the zero-$B$ value of
$\sigma^b_{xx}$, we finally obtain $n_b$ and $\mu_b$ separately for each value of $V_G$.
These are reported in Figs. \ref{figHall}b and \ref{figHall}c. 
The small values of $\mu_b$ (20-30 cm$^2$/Vs) result in a 
large $\mu_s/\mu_b\sim$ 100 and $\eta_H\sim$ 5. This
accounts for the pronounced low-$B$ curvatures seen in Fig. \ref{figHall}. 

The analysis implies high-mobility Dirac electrons in parallel with a much 
larger population of bulk electrons. Because of the 100-fold difference in mobilities,
the Dirac electrons produce 83$\%$ of the total weak-$B$ Hall conductance at large $|V_G|$. 
The fits include the surface Hall conductance from only one surface. Since its $G^s_{xy}$ 
already accounts for most of the observed $\sigma_{xy}$, there is very little room left
for a second surface term. We estimate that the Hall contribution from the other surface 
is less than 2$\%$ of $\sigma_{xy}$, which implies that 
its $\mu_s$ is $<$300 cm$^2$/Vs. This cannot produce resolvable SdH oscillations.

\section{Depletion-layer capacitance, screening and impurity band}
Our main results are on the tuning of the SdH oscillations of the Dirac surface states. However, the 
experiment also yields quantitative results on the electronic parameters in the depletion region,
which provide detailed picture of what happens under liquid gating.
A useful feature of the experiment is that, at each value of the applied gate voltage $V_G$, we can measure via the SdH oscillations
both $n_s$ and $E_F$ of the surface carriers
(hence the surface electrostatic potential $\varphi(0)$). In addition, we measure
the carrier density and conductivity of the bulk carriers, and the anion charge $Q$ accumulated on the
crystals surface. The 5 quantities provide a detailed picture of the
band-bending process as well as self-consistency checks in determining the depletion capacitance. 
We apply the standard analysis of field-effect gating~\cite{Stern,Ashcroft,Sze},
which is summarized in Appendix \ref{appcap}.

For gating to induce band bending, the chemical potential must already lie inside the bulk gap in zero $V_G$
(the case for Bi$_2$Te$_2$Se). [If, instead, $E_F$ lies high in the conduction band (as the case in as-grown Bi$_2$Se$_3$), the applied $E$-field
leads to Thomas Fermi screening~\cite{Ashcroft} for which the screening length is
$\lambda_{TF} = \sqrt{\pi a_B/4k_F}$ is typically a few $\rm \AA$ ($a_B = \hbar^2/me^2$ is the Bohr radius)].
For a hard gap (impurity band absent), a negative $V_G$ leads to a depletion region.
However, despite displaying a very large
bulk resistivity (2-6 $\Omega$cm) at 4 K, the current generation of Bi$_2$Te$_2$Se crystals still have a substantial
bulk carrier density ($n_b\sim$ 2$\times 10^{17}$ cm$^{-3}$). This implies an impurity band extends across the gap.
Nonetheless, band bending over a significant depletion region ($\sim$10 $\mu$m) is observed.
We will analyze this situation at the end of this section after we estimate the depletion capacitance (see also Appendix \ref{apptemp}).

For $V_G<0$, the $E$-field from the anions repels bulk electrons away from the surface,
exposing the ionized donors within the depletion width $d$. 
Figure \ref{figGate}a shows a sketch of the band bending near the surface exposed to the liquid.
For finite $V_G$, the ionic liquid polarizes to form, in effect, two capacitors each with
spacing of the order of the molecular radius $a$. Each capacitor stores the charge $Q$. 
The capacitor at the gate electrode 
has an area $A'$ much larger than that of the capacitor at the crystal surface $A$, so that
most of the potential drop $V_G-V_s$ falls across the latter ($V_s$ is the voltage 
corresponding to $\varphi(0)$ and the ground is taken deep in the bulk at $x\to +\infty$).
In Sample 1, $A$ = 2.9 mm$^2$ and $A'$ = 30 mm$^2$. 
The $E$-field produced by the anion layer just to the left of the crystal surface is 
$E(0^-) = Q/A\epsilon_0$.

\begin{figure}[t]
\includegraphics[width=7 cm]{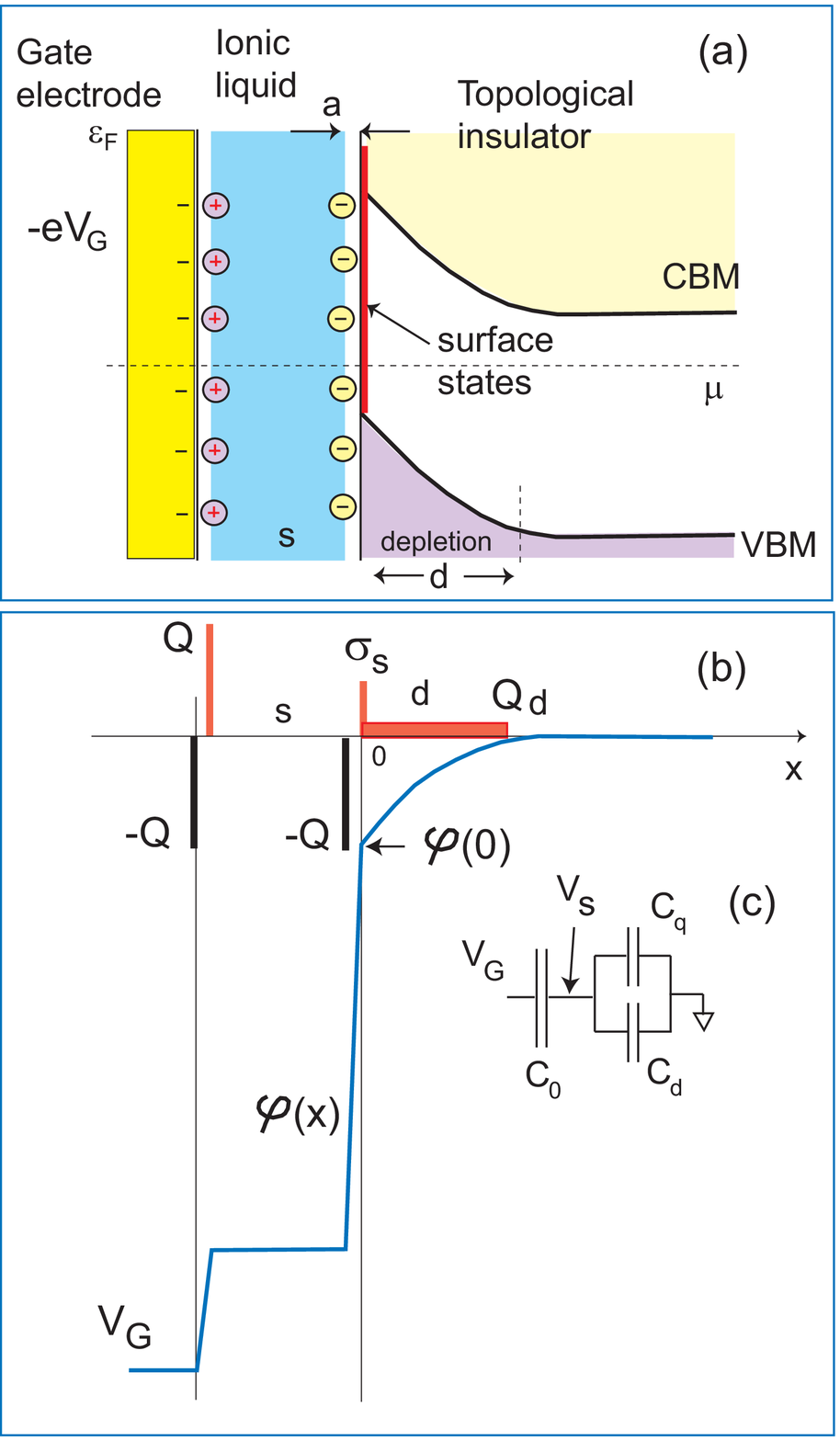}
\caption{\label{figGate} (color online)
Sketch of band bending and the profiles of $\rho(x)$ and $\varphi(x)$ in the liquid gating experiment.
Panel (a) shows upwards bending of the bands induced by a negative gate voltage $V_G$. The cations and anions 
define two series capacitors with spacing $a$ (molecular radius). The depletion layer in the bulk of the TI extends
a distance $d$. Panel (b) displays the charge distribution versus $x$. The negative charge $-Q$ on the gate electrode is
replicated by the anion layer separated by $a$ from the TI surface. This is compensated by the sum of the surface charge density
$\sigma_s$ and the ionized impurity charges inside the depletion layer (Eq. \ref{Q}). The electric potential $\varphi(x)$ corresponding to
$\rho(x)$ is sketched. Panel (c) shows the circuit of the equivalent capacitors $C_0$, $C_d$ and $C_q$.
}
\efig

In Fig. \ref{figGate}b, we sketch the profiles of the charge density $\rho(x)$ and the 
electrostatic potential $\varphi(x)$ (the $x$-axis is normal to the surface). 
Within the liquid, $\rho(x)$ of the cations and anions are taken to 
be delta functions of strength $\pm Q$. In the TI, the surface charge density is represented by
a delta function ($\sigma_s$). Within the bulk, however, $\rho(x)$ is distributed 
over the depletion layer to a depth $d$. As a guide, it is convenient to adopt the usual approximation, 
whereby $\rho(x)$ is taken to be uniform for $0<x<d$. 
In the uniform-charge approximation, $\varphi(x)$ varies as $-(x-d)^2$ in the depletion region.
Its value at the surface is then
\be
\varphi(0) = -N_ded^2/(2\epsilon_0\epsilon_s),
\label{phi0}
\ee
where $N_d$ is the donor impurity concentration and $\epsilon_s$ the screening dielectric parameter.
The charge $Q_d$ induced in the depletion width by $\varphi(0)$ defines the depletion capacitance
$C_d = N_dedA/\varphi(0) = \epsilon_0\epsilon_s A/d$. The surface charge density $\sigma_s$ 
induced by $\varphi(0)$ is represented by the quantum capacitance
$C_q = \sigma_s/\varphi(0) = e^2(dn_s/d\mu)$. Clearly, $C_d$ and $C_q$ are in parallel
combination (Fig. \ref{figGate}c).

The large slope change at $x=0$ mainly reflects the strong dielectric screening in the bulk of the
TI ($\sigma_s$ makes a negligible contribution). Thus the intense $E$-field produced by the
anions is strongly screened by polarization effects inside the crystal ($E(0^-)\gg E(0^+)$). 
As shown in Fig. \ref{figGate}c, the parallel combination of $C_d$ and $C_q$ is in 
series with $C_0$, the series combination of the cation and anion capacitances.
In all samples, we find that $C_d\gg C_q$, so we may ignore the quantum capacitance in the discussion below.

\noindent
\emph{Magnitude of $C_d$}\\
As shown in Fig. \ref{figError}b, $E_F$ in Sample 1 decreases by $\sim$100 mV when the applied $V_G$ is -2 V.
Thus, only a small fraction ($\sim 1/20$) of the applied gate voltage is effective in bending the band ($V_s$ = -0.1 V).
We can use this observation to determine the depletion capacitance $C_d$.
The value of $C_0/A$ for ionic liquids is 11-12 $\mu$F/cm$^2$~\cite{Koch}.
(From the expression, $C_0/A = \epsilon_0\epsilon_{liq}/a$, this corresponds to $\epsilon_{liq}$ = 4, and $a$ = 3 $\rm\AA$.) 
Using the ratio $V_s/(V_G-V_s)\sim C_0/C_d$ (neglecting $C_q$), we estimate that $C_d/A\simeq$ 240 $\mu$F/cm$^2$.

Alternatively, we may estimate $C_d$ by integrating the ionic current to find the charge $Q$.
For Sample 1 with $V_G$ = -2 V, the ionic charge current deposits a total negative ionic charge at the surface equal to
$-Q/A\sim 2\times 10^{14}e$/cm$^2$ = -3.2$\times 10^{-5}$ C/cm$^2$. 
Since $Q/A$ is stored in $C_d$ by the voltage $V_s\sim$ 0.1 V,
we have $C_d/A\sim$ 320 $\mu$F/cm$^2$, which is 33$\%$ larger than the first estimate, but within our uncertainties.
The main source of uncertainty is the actual area coated by the anions. Because the ions can coat the silver paint
contacts and voltage and current leads, the area can exceeds that of the crystal $A$ by 50 to 100$\%$.

[By equating $Q/A$ to $N_ded$ (see Eq. \ref{Q}), we may estimate the depletion width $d$ as a check.
The donor density $N_d$ is roughly equal to the bulk density observed at 4 K, $n_b~\sim 2\times 10^{17}$ cm$^{-3}$.
This gives $d\simeq$ 10 $\mu$m. The deep penetration of the depletion region into the bulk
is consistent (within a factor of 2) with the 40$\%$ change observed in the resistivity and Hall coefficient at 4 K.]

Taking the range $C_d/A$ = 240-320 $\mu$F/cm$^2$, we find that 
the depletion capacitance is 5,000-6,000$\times$ larger
than the values commonly observed in a Si-MOSFET device ($C_{d,Si}/A\simeq$ 0.05-0.06 $\mu$F/cm$^2$~\cite{Stern,Sze}).
The enhancement points to a very large polarizability in the ground state of Bi$_2$Te$_2$Se
when $E_F$ lies inside the energy gap. This is perhaps unsurprising given that the 
energy gap in high-purity Si is devoid of impurity states. By contrast, 
Bi$_2$Te$_2$Se at 4 K displays a small, but metallic conductivity arising from a large population of impurity-band electrons.

Here, we resume discussion of the finite DOS in the gap.
To create an extended depletion region with significant band bending, as we have here (Fig. \ref{figGate}a), 
the weak bulk conductivity must be further driven to zero throughout the depletion region
in order to sustain a finite $E$-field (otherwise one has Thomas-Fermi screening with the very short screening length 
$\lambda_{TF} \simeq$ 6 $\rm\AA$
for $n_b$ = 2$\times 10^{17}$ cm$^{-3}$m). To explain how band-bending is sustained over a large depletion region,
we need the existence of a mobility edge in the 
impurity band. Throughout the depletion layer, $E_F$ lies below the moblity edge so that the
conductivity is vanishingly small at 4 K. Because impurity-band states close to the mobility edge 
have a greatly enhanced polarizability in an $E$-field, we expect the electronic contribution to 
dielectric constant $\epsilon_s$ to be orders of magnitude larger than the lattice contribution.
Measurements of $C_d/A$ probe directly the electronic polarizability in the depletion region. A
possible scenario is described in Appendix \ref{appcap} and Fig. \ref{figimp}.

\section{Conclusions}
Applying the relatively new technique of ionic liquid gating to bulk crystals of Bi$_2$Te$_2$Se with
resistivity exceeding 4 $\Omega$cm at 5 K, we find that $E_F$ of the surface Dirac fermions can be
tuned over a considerable range.
In contrast to previous gating experiments, we readily resolve the surface SdH oscillations 
at each value of the gate voltage. By measuring the SdH period, we find that the surface Fermi energy $E_F$
(Sample 1) decreases  
from 180 mV to 75 mV relative to the Dirac Point as $V_G$ is changed from 0 to -2.8 V. 
In a field of 14 T, the lower limit corresponds to
the middle of the broadened $N$ = 1 Landau Level. Attaining such low Landau levels enables the $-\frac12$
intercept (predicted for Dirac fermions) to be determined with high accuracy.
We also find that the intercepts are closely similar for a broad range of $V_G$ in both Samples 1 and 2.

Using the surface mobility $\mu_s$ and density $n_s$ determined from the SdH oscillations,
we find that the Dirac fermion Hall conductivity from the surface exposed to the anions accounts
for up to 83$\%$ of the total observed weak-$B$ Hall conductivity at 5 K. The analysis allows an accurate determination
of the bulk carrier mobility and density at each $V_G$ (Fig. \ref{figHall}). The picture inferred is that,
with gating, the density $n_s$ of the surface Dirac fermions decreases steeply while their mobility
$\mu_s$ increases to a maximum value of 2,400 cm$^2$/Vs. The bulk carriers are depleted to a depth of 10 $\mu$m from the 
surface, with $\mu_b$ remaining at the low value 20 cm$^2$/Vs.

The large enhancement of $\mu_s$
by liquid gating (Fig. \ref{figHall}b) is perhaps the most intriguing 
feature reported here. To our knowledge, this is the first realization of 
enhancement of surface SdH amplitudes by an \emph{in situ} technique.~\cite{mobility}
A recent STM experiment~\cite{Beidenkopf} reveals that
the Dirac Point closely follows spatial fluctuations of the local
potential on length scales of 30-50 nm. This could lead to strong scattering
of surface electrons. We speculate that, under liquid gating, the anions accumulate
at local maxima in the potential, thereby levelling out the strongest spatial fluctuations.
The results provide encouragement that alternative routes that even out local
potential fluctuations can lead to further improvements in $\mu_s$.

To address the question whether ionic liquid gating actually alters the carrier concentration by chemical
reaction (as opposed to simply bending the band), we have performed extensive tests to separate the two effects.
By carefully selecting the experimental conditions (e.g. the gating temperature),
monitoring charge accumulated $Q$, and checking for reversibility, we establish that band-bending is the dominant effect
in these experiments. Lastly, the 5 quantities measured at each gate voltage setting ($E_F$, $n_s$, $n_b$, $\rho$ and $Q$)
provide a quantitative picture of the gating process. The depletion
capacitance measured implies that, within the depletion region, the electronic polarizability is strongly enhanced.

\vspace{3mm}
{\bf Acknowledgements} 
We thank Joe Checkelsky, Jianting Ye, and Hongtao Yuan for advice on liquid gating.
The research is supported by the Army Research Office (ARO W911NF-11-
1-0379) and by the US National Science Foundation (Grant No. DMR
0819860). Sample growth and characterization were supported
by an award from the Defense Advanced Research Projects
Agency under SPAWAR Grant No. N66001-11-1-4110. High-field measurements were
performed at the National High Magnetic Field Laboratory
which is supported by NSF (Award DMR-084173), by the State of Florida, and
by the Department of Energy.


\renewcommand{\thefigure}{A\arabic{figure}}
\renewcommand{\thesection}{A\arabic{section}}
\renewcommand{\theequation}{A\arabic{equation}}
\setcounter{section}{0}
\setcounter{figure}{0}
\setcounter{equation}{0}

\section{{\bf Appendix I: Gating temperature}}\label{apptemp}

\begin{figure}[t]
\includegraphics[width=9 cm]{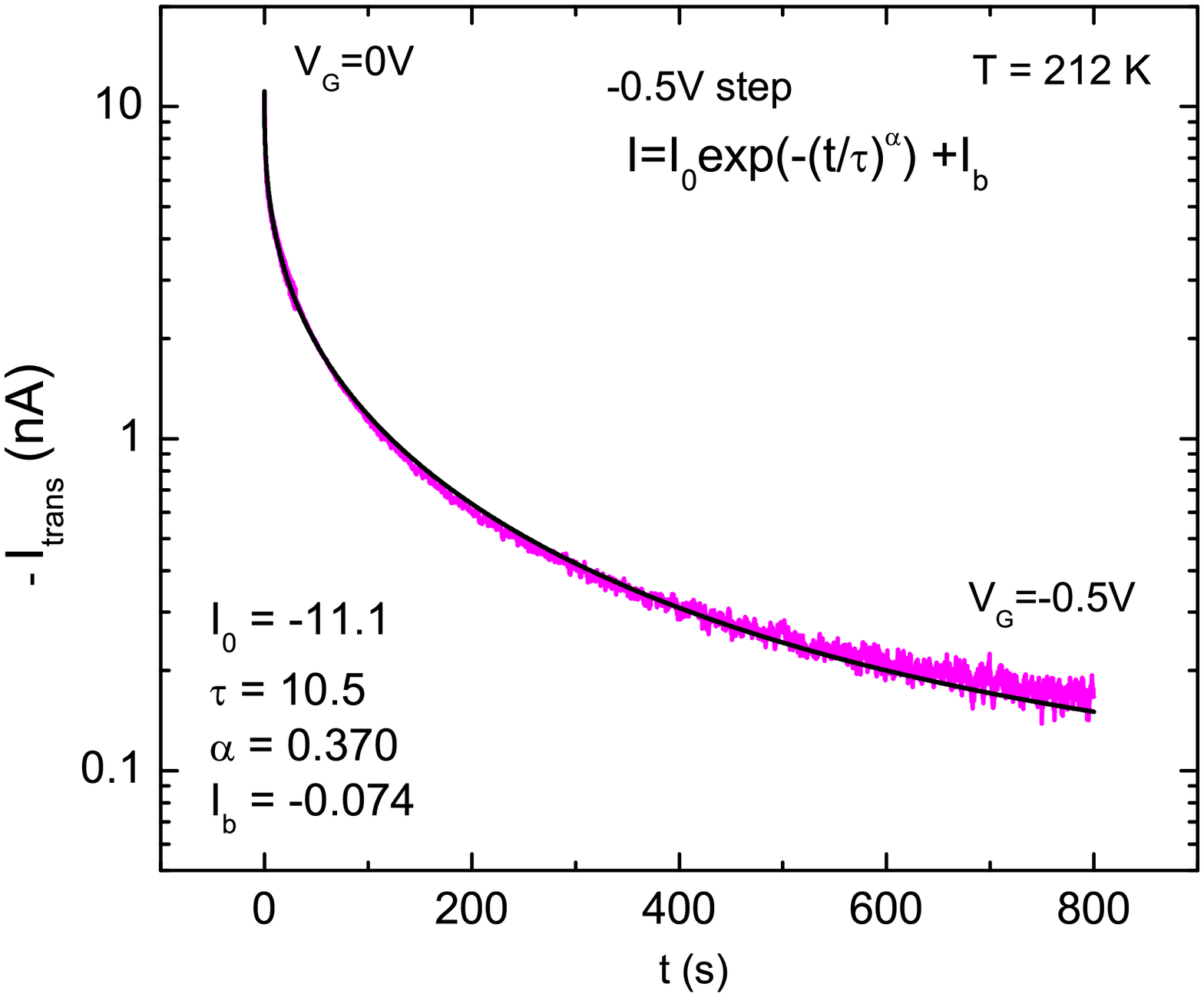}
\caption{\label{figtrans} (color online) 
The transient (discharging) current $I_{trans}$ versus time $t$ following a step-change of
$V_G$ from 0 to -5 V at $t$ = 0 at $T$ = 212 K in Sample 3. The observed current fits well to the stretched exponential form
Eq. \ref{Itrans} with the parameters $I_0$ = -11.1 nA, $I_b$ = -0.074 nA, $\alpha$ = 0.37, and $\tau$ = 10.5 s.
$I_b$ is the long-term steady-state background current. 
}
\efig

\begin{figure}[t]
\includegraphics[width=9 cm]{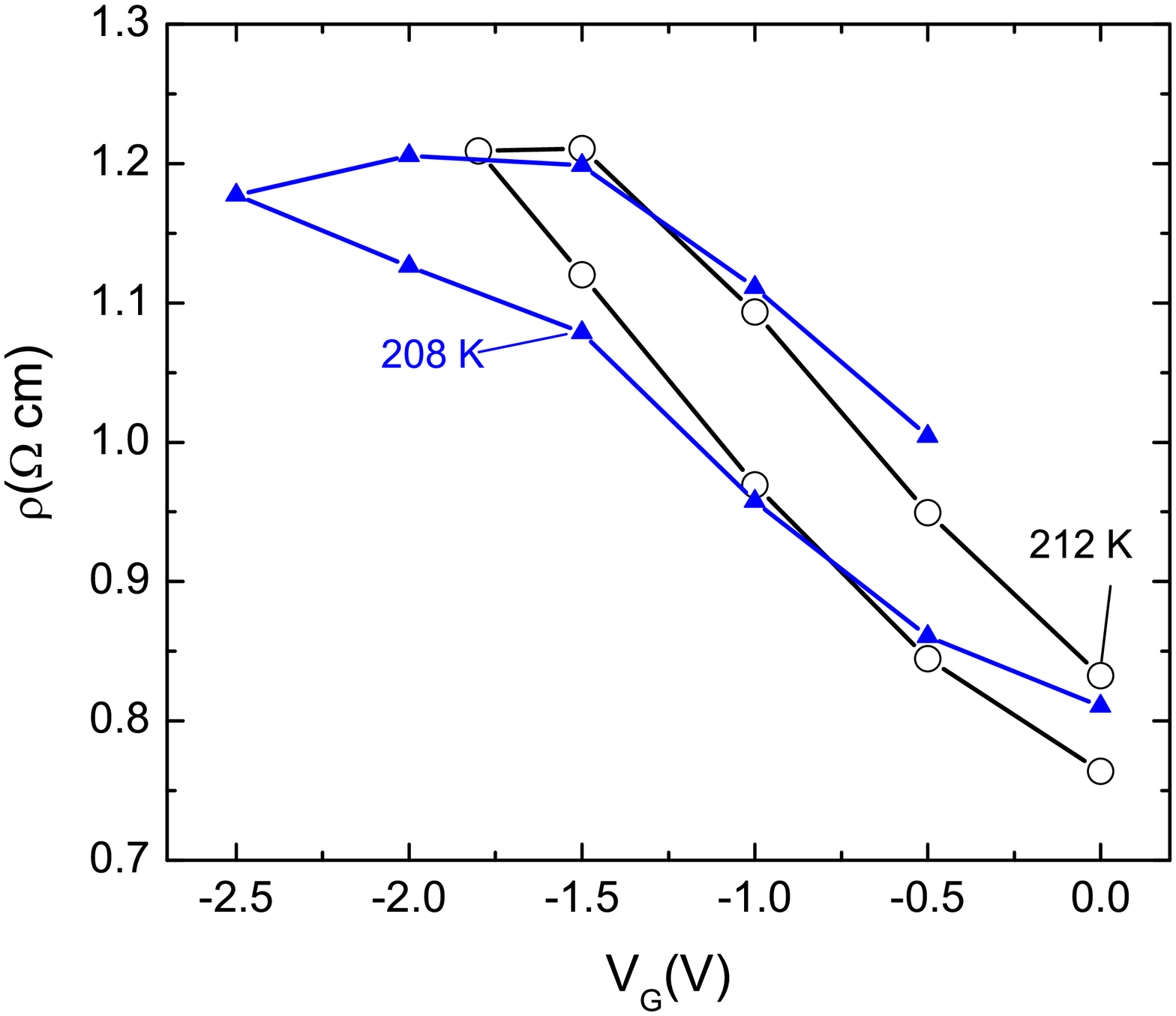}
\caption{\label{figRLowT} (color online) 
Apparent hysteretic behavior of $\rho$ vs. $V_G$ observed at temperatures close to the glass transition $T_G$.
Varying $V_G$ at $T$ too closer to $T_G$ strongly suppresses the background quasi-steady state $I_1$ and 
possibilities of chemical reaction. However, this imparts increased hysteresis in $\rho$ when $V_G$
is cycled (here $T$ = 208 and 212 K). The Hall density $n_H$ shows similarly large hysteresis (not shown). 
As discussed in the text, we show that this apparent low-temperature hysteresis results from incomplete
melting of the ionic liquid.
}
\efig

\begin{figure}[t]
\includegraphics[width=9 cm]{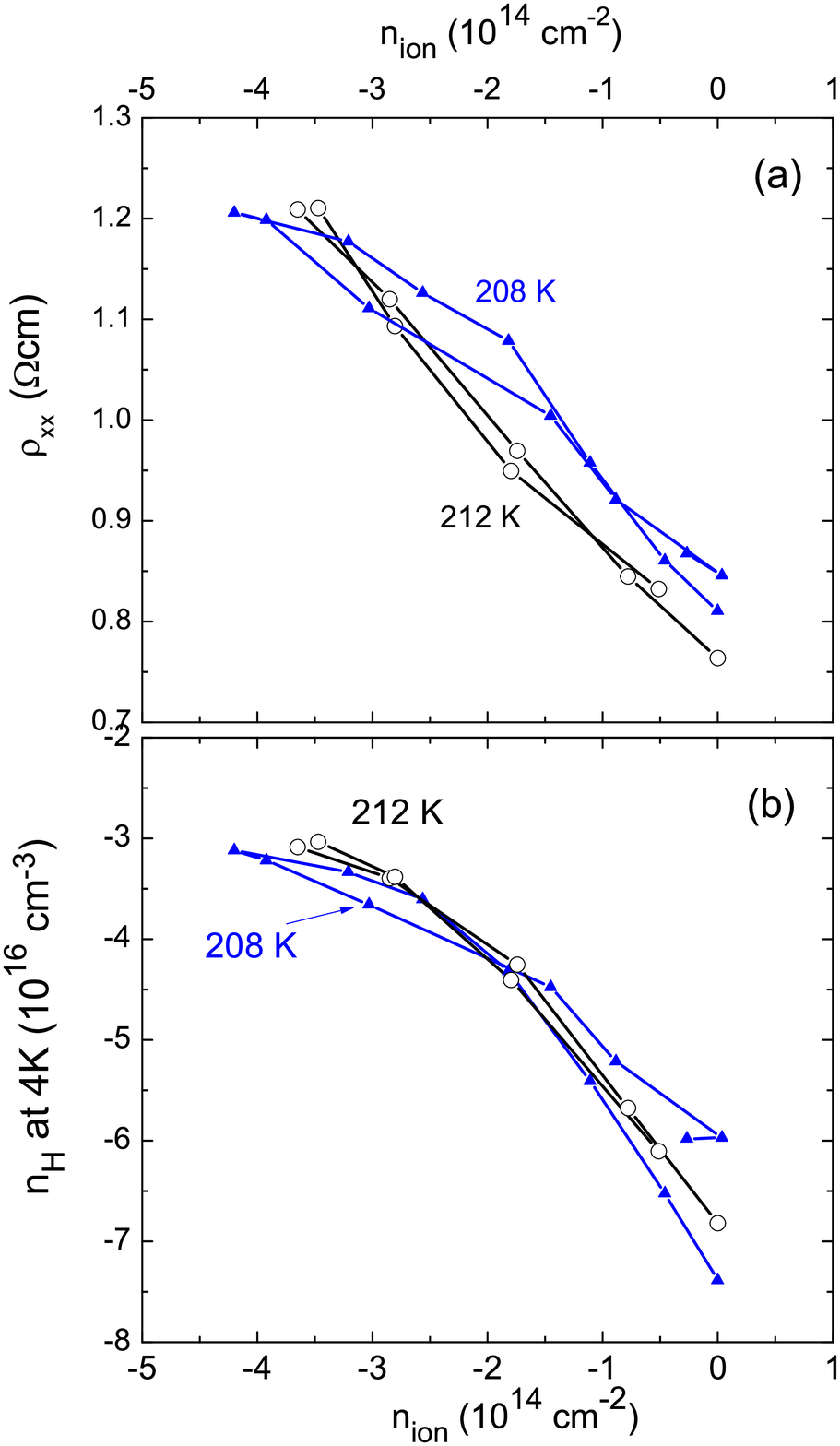}
\caption{\label{figNion} (color online) 
Absence of low-temperature hysteresis when $\rho$ and $n_H$ are plotted against $n_{ion}=Q/eA$, with $A$ = 2.9 mm$^2$.
Replotting the data for $\rho$ in Fig. \ref{figRLowT} versus $n_{ion}$ (instead of $V_G$) removes the hysteresis apparent in Fig. \ref{figRLowT}.
This shows that the hysteretic behavior arises from the variation of $Q$ vs. $V_G$. The physically relevant quantities inside the
crystal $\rho$ and $n_H$ are dependent only on $Q$, strongly supporting the premise that band-bending 
produces these changes rather than chemical reaction.}
\efig

We discuss here the factors that dictate our choice of the ``gating'' temperature.
Insight on the ion accumulation process is provided by monitoring the transient current
$I_{trans}$ following a step-change in $V_G$ (with $T$ fixed in the interval 208 $<T<$ 260 K).
As the ions flow to adjust to the new potential, $I_{trans}(t)$ decays over a time scale of 10$^3$ s. 
As shown in Fig. \ref{figtrans}, $I_{trans}(t)$ fits well to the stretched exponential form
\be
I_{trans}(t) = I_0{\rm e}^{-(t/\tau)^\alpha} + I_b,
\label{Itrans}
\ee
where $\alpha$ varies from 0.35 to 0.50 depending on $T$ and $I_b$ is the long-term, steady-state
background current. Integrating the transient part, we obtain the ionic charge accumulated at time $t$,
$Q(t) = \int^{t}_0 dt' [I_{trans}(t') - I_b]$.

In our experiment, the optimal temperature falls in the window
220-240 K. As shown in Fig. \ref{figtrans}, the background $I_b$ is a factor of 10-100 smaller than
the onset value $I_0$. Raising $T$ above 240 K leads to an exponential increase in $I_b$. 
Most of this arises from the finite (if small), thermally-activated bulk conductivity of the ionic liquid.
In addition, chemical reaction adds an increasingly important component to $I_b$ when
$|V_G|$ exceeds 2 V. For these reasons, we keep $T$ under 240 K.

To minimize possible chemical reaction, it might seem expedient to lower the gating $T$ to as close to 
the glass transition as feasible (this strongly suppresses $I_b$). 
However, we quickly encounter a different problem, namely the failure of
the ionic charge configuration to melt completely. As a result, $Q(t)$ fails to attain
its equilibrium value as $V_G$ is changed (even if $t\gg 10^3$ s), leading to a different kind of hysteresis.

Figure \ref{figRLowT} plots the changes in $\rho$ (Sample 3) as  $V_G$ is cycled between 0 and -2.5 V at 
the relatively low $T$ (208 and 212 K). In contrast to Fig. \ref{figtest}, we observe
sizeable hysteretic behavior (also in $n_H$, not shown). To show that this is not caused by
chemical reaction (which should be greatly suppressed at low $T$), we have measured the accumulated
$Q(t)$ and found that it displays the same hysteresis (vs. $V_G$). When we plot the changes
in $\rho$ and $n_H$ against $n_{ion} = Q/eA$ (see Fig. \ref{figNion}), 
the hysteretic behavior apparent in Fig. \ref{figRLowT} is largely removed.

This implies that, at these low $T$, a significant portion of the ionic ``solid'' accumulated at the
previous value of $V_G$ fails to melt and flow in response to the new $V_G$. Hence
$Q(t)$ never attains its equilibrium value even at long $t$. This leads 
to strong hysteresis in $Q$ vs. $V_G$. However,
the near-absence of hysteresis in Fig. \ref{figNion} shows that $\rho$ and $n_H$ adjust reversibly to
the non-equilibrium value of $Q$. The key parameter that
causes $\rho$ and $n_H$ to change is the electric-field $E(0^-)$ produced by $Q$ even when it lags
the applied $V_G$. This direct link provides further support for our conclusion that
the dominant effect of changing $Q$ is band-bending.

\section{{\bf Appendix II: Depletion and quantum capacitances}}\label{appcap}

\begin{figure}[t]
\includegraphics[width=7.5 cm]{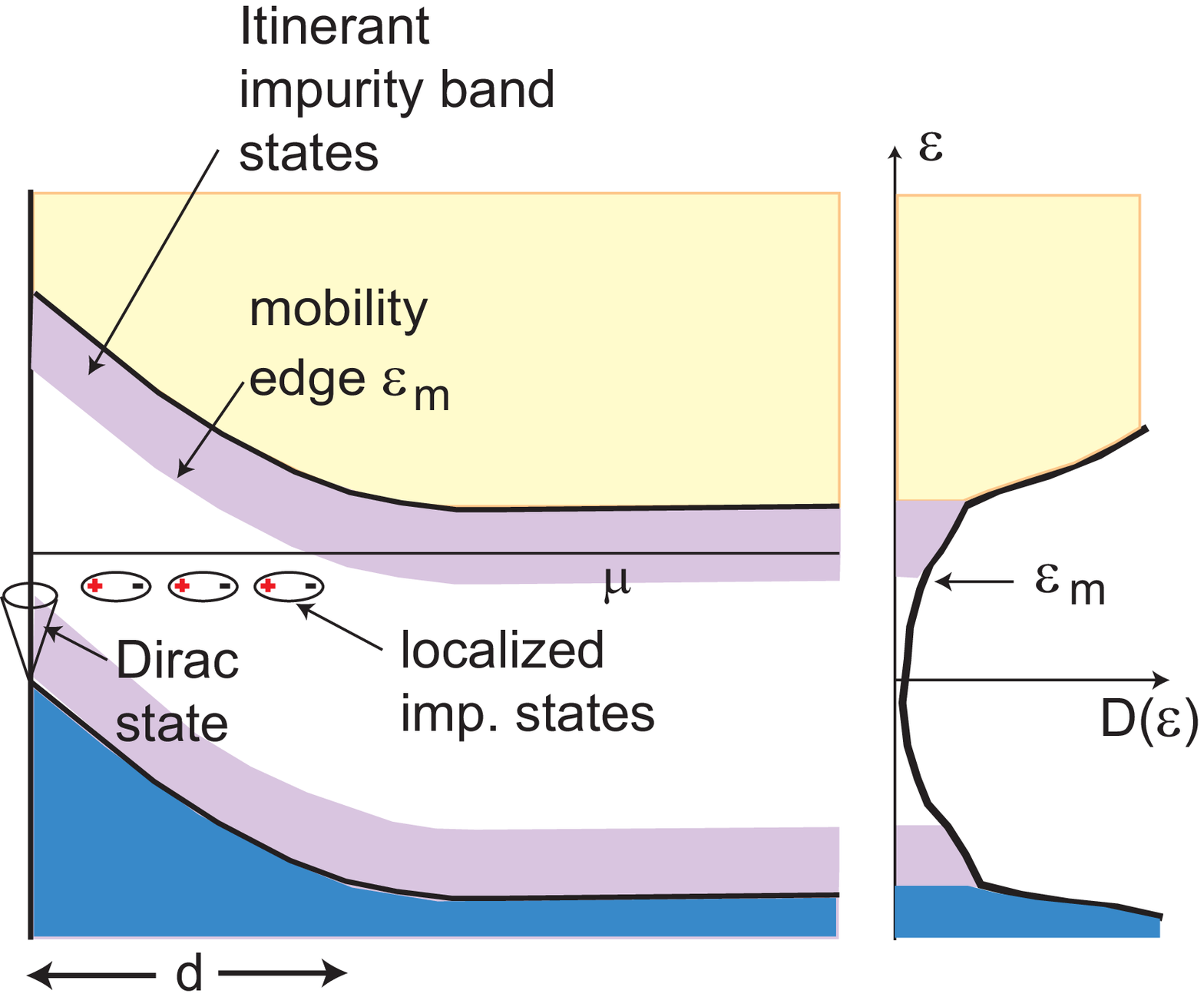}
\caption{\label{figimp} (color online) 
Model for the impurity band states implied by the liquid-gating experiment. The density of states (DOS)
profile ${\cal D}(\varepsilon)$ across the bulk energy gap is sketched on the right. The impurity-band DOS tapers deep
into the gap, as suggested by STM experiments~\cite{Beidenkopf}. Impurity states lying above the mobility edge $\varepsilon_m$ (shaded)
are itinerant with mobility $\mu_b\sim$ 20 cm$^2$/Vs. States below $\varepsilon_m$ are strongly localized at 4 K. 
Near the surface (left sketch), $\varepsilon_m$ is lifted above 
$\mu$ within the depletion layer, substantially decreasing the bulk contribution to the observed $\sigma$ and $\sigma_H$. 
Localized states near the mobility edge (sketched as dipoles) contribute strongly to dielectric screening because of their 
enhanced polarizability.
}
\efig

With reference to Fig. \ref{figGate}, the free-charge density profile $\rho(x)$ is comprised of 
4 delta functions $\delta(x)$ and an extended distribution over the depletion width $d$ (which we assume has a flat 
profile expressed by the step-function $\theta(x)$~\cite{Ashcroft}, as shown in Fig. \ref{figGate}b). Setting the origin $x=0$ at the TI surface, we have
\begin{eqnarray}
\rho(x) &=& -\frac{Q}{A'}\delta(x+s) + \frac{Q}{A'}\delta(x+s-a) -\frac{Q}{A}\delta(x+a) \notag\\
 				&& +\sigma_s \delta(x) + N_ded\,[\theta(x)-\theta(x-d)],
\label{rho}
\end{eqnarray}
where $\sigma_s$ is the surface charge density at the exposed crystal face,
and $N_d$ is the density of ionized donor impurities within the depletion width $d$ (we take $e>0$).
The electrostatic potential $\varphi(x)$ is derived from the Poisson equation
\be
-\varepsilon(x) \frac{\partial^2\varphi}{\partial x^2} = \frac{\rho(x)}{\epsilon_0},
\label{phi}
\ee
with $\epsilon_0$ the vacuum permittivity. The dielectric function $\varepsilon(x) = \epsilon_s$ inside the TI ($x>0$). Within the ionic liquid,
$\varepsilon(x) = \epsilon_{liq}$.

Integration of Eq. \ref{phi} gives the profile of $\varphi(x)$ sketched in Fig. \ref{figGate}b. We wish to relate the
charge density $Q/A$ to $\sigma_s$ and $d$.
Setting $\varphi$ and $\partial\varphi/\partial x$ to 0 
deep in the bulk ($x>d$), we have for the $E$-field just to the right of $x = 0$
\be
E(0^+) = -\left(\frac{\partial\varphi}{\partial x}\right)_{0+} = - \frac{N_ded}{\epsilon_0\epsilon_s}.
\label{E0}
\ee
In the flat-profile approximation for $\rho(x)$, Eq. \ref{phi} gives the parabolic variation of $\varphi(x)$ 
\be
\varphi(x) = -\frac{N_de}{2\epsilon_0\epsilon_s}(x-d)^2   \quad (x>0).
\label{varphi}
\ee

Next, we integrate Eq. \ref{phi} between the limits $x = 0^{\pm}$ (bracketing $x =0$) to get
\be
\epsilon_s E(0^+) - E(0^-) = \sigma_s/\epsilon_0.
\label{EE}
\ee
Together, Eqs. \ref{E0} and \ref{EE} give for the $E$-field just to the left of the surface
\be
E(0^-) = -(\sigma_s + N_ded)/\epsilon_0.
\label{E0-}
\ee
This strong $E$-field emanating from the anion charge $-Q$ is only partially screened by the surface charge $\sigma_s$. The remaining
$E$ penetrates a distance $d$ into the bulk until screened by enough ionized donor charge. The lattice polarizability,
expressed by the bulk dielectric constant $\epsilon_s$,
also contributes to the screening. (It is helpful to represent the dielectric screening, alternatively, as a 
bound-surface charge density $\sigma_b = -\epsilon_0(\epsilon_s-1)E(0^+)$ at $x$ = 0. However, this bound charge
should not be included in $\rho(x)$).

Finally, identifying $E(0^-)$ with the $E$-field within the molecular layer, $-Q/A\epsilon_0$, we arrive at the equation
\be
\frac{Q}{A} = N_ded + \sigma_s.
\label{Q}
\ee
We note that Eq. \ref{Q} is independent of $\epsilon_s$.
The charge $Q$ induced by the anions is partitioned between two charge reservoirs which see the same potential drop $V_s = \varphi(0)$ relative to the 
ground at $x=+\infty$. Hence, as shown in Fig. \ref{figGate}c, we regard the two charge reservoirs 
as two capacitors in parallel, namely the 
quantum capacitance~\cite{Luryi}
\be
C_q = \frac{\sigma_s}{\varphi(0)} = e^2\frac{dn_s}{d\mu},
\label{Cq}
\ee 
and the depletion-layer capacitance 
\be
C_d = N_dedA/\varphi(0).
\label{Cd}
\ee
Whereas in graphene, the quantum capacitance is readily resolved,
here it is shunted by the large $C_d$.

The parallel combination $C_q+C_d$ is in series with the ionic-liquid capacitor $C_0$ (the series combination of the cation and anion capacitors). The
voltage drop across $C_0$ is $V_G-V_s$.

\vspace{3mm}
\noindent
\emph{Mobility edge and electronic polarizability}\\
We discuss a scenario in which a strongly enhanced electronic polarizability arises 
within the depletion layer. Figure \ref{figimp} is a sketch of the band bending near the surface. 
As shown, the chemical potential $\mu$ in the bulk lies just below the bottom of the conduction band.
The right panel plots the density of states ${\cal D}(\varepsilon)$ in a cut in the bulk.
The impurity band is comprised of ``tails'' of ${\cal D}(\varepsilon)$ which taper downwards (upwards) from the
conduction band (valence band)~\cite{Beidenkopf}. At 4 K, the mobility edge
$\varepsilon_m$ sharply divides states that are itinerant (closer to the gap edge) from 
the states that are localized. Electrons in the itinerant states diffuse with the observed mobility $\mu_b\sim$ 20 cm$^2$/Vs. 
In an $E$-field strong enough to cause band bending, 
$\varepsilon_m$ is lifted above $\mu$ within the depletion region of width $d$. Occupied states within this 
region are strongly localized, so they do not contribute to the observed conductivity or Hall effect. 
However, because the localization length $\xi_{loc}$ diverges as $\varepsilon\to \varepsilon_m$ from below, 
the localized states have a greatly enhanced electronic polarizability. 
The electronic component of the dielectric screening parameter will be much
larger than that from the lattice polarizability.


\begin{thebibliography}{99}
\bibitem{mobility} 
Previous gating experiments 
reported either a decrease in $\mu_s$ with finite gating~\cite{Fuhrer},
or a non-monotonic variation of $\mu_s$ that trends downwards 
with increasing $|V_G|$~\cite{Ando12}.
SdH oscillations were not resolved in these studies.
\end{thebibliography}
\end{document}